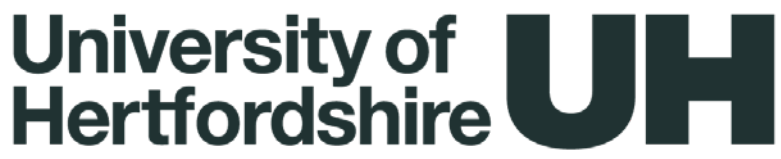

School of Physics, Engineering
and Computer Science

# Early Comparative Evaluation of Transformer Models for Multilingual Software Vulnerability Detection

**Fiza Naseer[1*], Javad Khan[1,] Muhammad Yaqoob[1], and Alexios Mylonas[1]**

*School of Physics, Engineering and Computer Science, University of Hertfordshire, Hatfield, UK*

Corresponding author: f.naseer@herts.ac.uk, j.a.khan@herts.ac.uk, m.yaqoob3@herts.ac.uk, a.mylonas@herts.ac.uk.

## Abstract

**Software vulnerability detection is increasingly important as modern applications combine multiple programming languages. This paper presents an early comparative evaluation of BERT, RoBERTa, and CodeBERT for binary vulnerability detection across HTML, Python, JavaScript, and PHP using the CVEFixes dataset and language-wise three-fold stratified cross-validation. The results show clear performance differences across languages, indicating that multilingual vulnerability detection requires more language-aware and robust transformer-based modelling strategies.**

## Introduction

Software vulnerabilities are a significant source of security incidents, financial damage, and increased software maintenance effort. Their impact has become more severe as software is increasingly deployed in critical sectors such as healthcare [2] and finance [8], where vulnerabilities can lead to privacy violations, service disruption, unauthorised access, and substantial economic loss [10]. The Common Weakness Enumeration (CWE) catalogue contains 943 weaknesses in version 4.17 [1], demonstrating the breadth of security issues in modern software systems.

To address these risks, researchers have proposed a range of vulnerability detection techniques, including static, dynamic, and hybrid analysis, as well as machine learning and deep learning approaches [9]. Traditional program-analysis techniques are often affected by high false-positive rates, limited code coverage, or poor practical efficiency [9]. Learning-based methods can improve automation and reduce manual effort; however, traditional machine learning approaches rely heavily on handcrafted features, while many deep learning methods still perform vulnerability prediction at a coarse-grained file or function level [11].

More recently, transformer-based models have shown promising results in the field of software vulnerability detection because of their ability to capture contextual information and long-range dependencies in source code [ 6, 7, 12, 14]. Many researchers have reported encouraging results using transformer architectures for vulnerability-related tasks [3, 14]. Nevertheless, most existing work focuses on one or only a few programming languages, for example C/C++, which limits applicability in multilingual software environments [5]. Since modern software systems increasingly combine multiple programming languages, effective vulnerability detection methods must generalise across different syntax, semantics, and coding patterns. Although several studies have explored multilingual settings, they remain limited in terms of language coverage and achieved performance, which calls for better generalisation [4]. Inspired by this gap, we present an early language-wise comparison of BERT, CodeBERT, and RoBERTa for vulnerability detection in HTML, Python, JavaScript, and PHP.

The contributions of this study are as follows:

- An early multilingual transformer-based pipeline for software vulnerability detection.
- A language-wise comparison of BERT, CodeBERT, and RoBERTa.

- Empirical evidence that transformer-model performance varies substantially across programming languages.

## Experimental/Simulation

The study views software vulnerability detection as a binary classification problem, as shown in Figure 1, where each source-code sample is labelled as vulnerable or safe. The CVEFixes dataset [13] contains source-code samples from multiple programming languages. This study focuses on HTML, Python, JavaScript, and PHP.

The selected subset of the CVEFixes dataset contains source-code samples from HTML, Python, JavaScript, and PHP. For this study, three fields were used: code, language, and safety. The safety field was treated as the binary class label, with vulnerable samples encoded as 1 and safe samples encoded as 0. After preprocessing, the final subset contained 285 HTML samples, including 143 vulnerable and 142 non-vulnerable samples (50.18% and 49.82%); 1,564 Python samples, including 782 vulnerable and 782 non-vulnerable samples (50.00% each); 1,562 JavaScript samples, including 781 vulnerable and 781 non-vulnerable samples (50.00% each); and 5,589 PHP samples, including 2,795 vulnerable and 2,794 non-vulnerable samples (50.01% and 49.99%). This near-balanced distribution reflects the selected experimental subset used in this study and may not represent the complete class distribution of the full CVEFixes dataset.

HTML, Python, JavaScript, and PHP were selected because they are commonly used in web-based and server-side software systems. JavaScript and PHP are frequently used in dynamic web applications. Python is widely used for backend services, automation, and data-driven applications, while HTML appears in web-facing and template-related code. The selected languages also differ in syntax, structure, and typical vulnerability patterns, which makes them suitable for an early language-wise comparison. However, they do not represent the full range of programming languages. Therefore, the findings should be interpreted as an early multilingual evaluation rather than a fully generalisable multilingual vulnerability detection benchmark.

During preprocessing, incomplete records and missing or empty code samples were removed. Only the code, language, and safety fields were retained for the final experiments. Source-code samples were tokenised using a maximum sequence length of 512 tokens. BERT, RoBERTa, and CodeBERT were evaluated with their corresponding tokenisers. Training was performed using language-wise three-fold stratified cross-validation to preserve class distribution across folds. The main hyperparameters included learning rate, batch size, number of epochs, and weight decay. Fine-tuning was implemented using the Hugging Face Trainer framework, and performance was measured using accuracy, precision, recall, and F1-score.

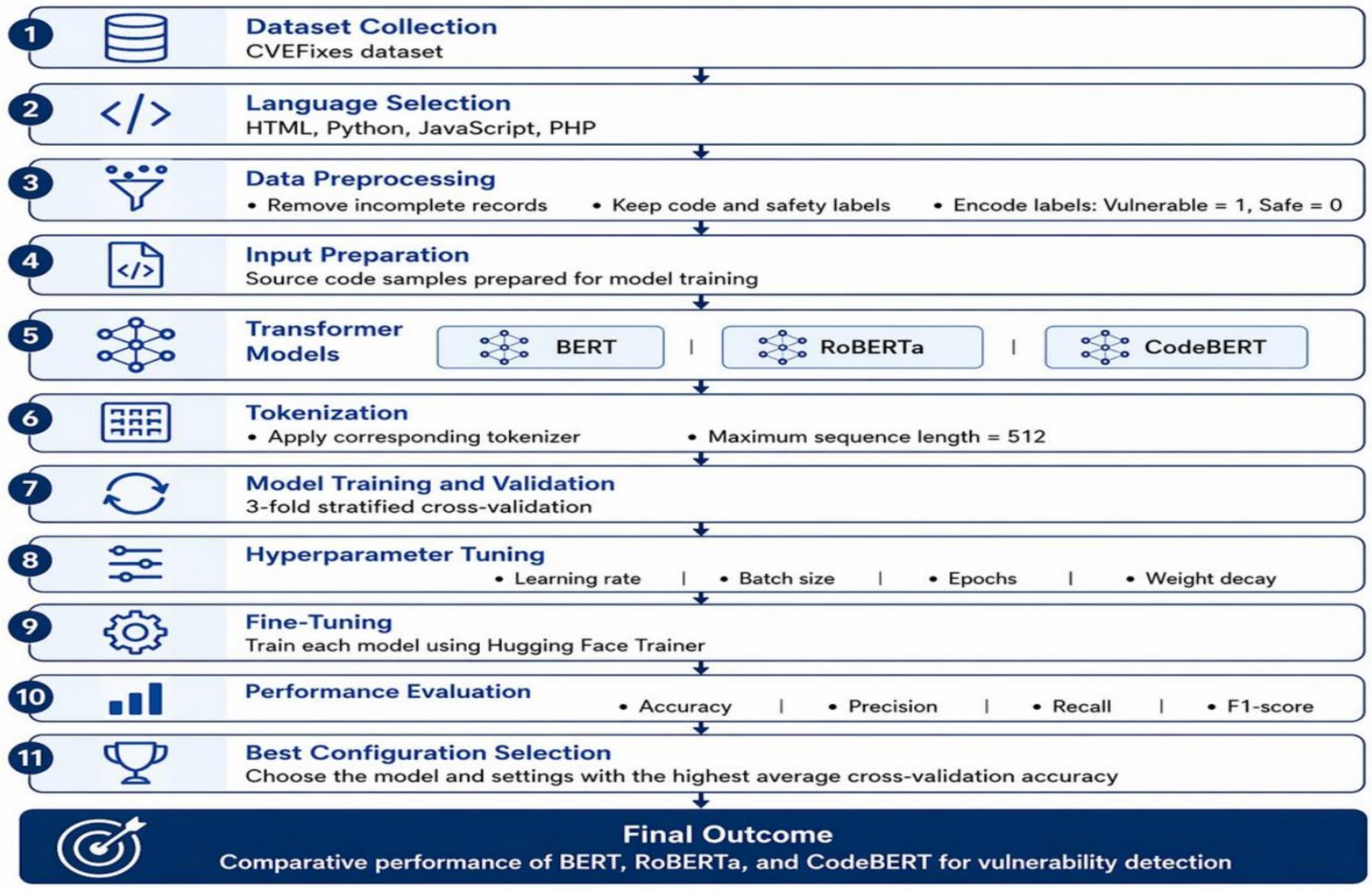

*Figure 1. Methodology Diagram*

## Results and Discussion

Table 1 summarises the early language-wise results for BERT, RoBERTa, and CodeBERT. The results show that transformer performance varies substantially across programming languages. The strongest overall performance was obtained for HTML, where CodeBERT achieved an F1-score of 0.7171. BERT and RoBERTa also performed relatively well on HTML, with F1-scores above 0.70. In contrast, Python, JavaScript, and PHP produced lower F1-scores, all below 0.45, suggesting that these languages remain more challenging for the evaluated models in the current experimental setting.

*Table 1. Early language-wise results for BERT, RoBERTa, and CodeBERT.*

| Language | Model | Accuracy | Precision | Recall | F1-score |
|---|---|---|---|---|---|
| HTML | BERT | 0.7208 | 0.7673 | 0.7208 | 0.7064 |
| HTML | RoBERTa | 0.7246 | 0.7854 | 0.7246 | 0.7078 |
| HTML | CodeBERT | 0.7359 | 0.8208 | 0.7359 | 0.7171 |
| Python | BERT | 0.5473 | 0.7563 | 0.5473 | 0.4314 |
| Python | RoBERTa | 0.5473 | 0.7454 | 0.5473 | 0.4330 |
| Python | CodeBERT | 0.5480 | 0.7293 | 0.5480 | 0.4367 |
| JavaScript | BERT | 0.5333 | 0.7354 | 0.5333 | 0.4060 |
| JavaScript | RoBERTa | 0.5397 | 0.7042 | 0.5397 | 0.4289 |
| JavaScript | CodeBERT | 0.5352 | 0.7044 | 0.5352 | 0.4177 |
| PHP | BERT | 0.5212 | 0.6379 | 0.5212 | 0.4036 |
| PHP | RoBERTa | 0.5182 | 0.7501 | 0.5182 | 0.3728 |
| PHP | CodeBERT | 0.5196 | 0.7498 | 0.5196 | 0.3758 |

The results indicate that no single model consistently outperforms the others across all languages. CodeBERT achieved the best F1-score for HTML and Python, RoBERTa achieved the highest F1-score for JavaScript, and BERT achieved the strongest F1-score for PHP. This variation suggests that vulnerability patterns are language dependent and that multilingual vulnerability detection may require language-specific tuning, improved class-imbalance handling, and richer source-code representations. These findings also support the need for future studies using broader datasets and more robust evaluation measures, including vulnerable-class recall, macro F1-score, Matthews correlation coefficient, and PR-AUC.

## Conclusion

This paper presented an early comparative evaluation of BERT, RoBERTa, and CodeBERT for multilingual software vulnerability detection across HTML, Python, JavaScript, and PHP. The findings show that transformer-based models can support binary vulnerability detection, but their performance varies considerably across programming languages. The best results were achieved on HTML, while Python, JavaScript, and PHP remained more challenging in the current experimental setting. These differences may be influenced by language syntax, coding patterns, dataset quality, and class distribution. Overall, the results suggest that transformer models may not generalise equally well across all programming languages. Future work will expand the language coverage, investigate mixed-language and language-specific training strategies, apply stronger imbalance-handling techniques, and evaluate additional transformer-based models using more robust vulnerability-focused metrics.